\documentclass[preprint]{emulateapj}
 
\slugcomment{To appear in {\it{The Astrophysical Journal}}}
\shortauthors{Gonzalez et al.}
\shorttitle{A Multiply Imaged LIRG Behind the Bullet Cluster}

\newcommand{\spitzer}{{\it Spitzer}}
\newcommand{\hst}{{\it HST}}
\newcommand{\chandra}{{\it Chandra}}
\newcommand{\irac}{IRAC}
\newcommand{\mips}{MIPS}
\newcommand{\acs}{ACS}
\newcommand\kcorrect{\texttt{kcorrect}}
\newcommand\multidrizzle{\texttt{MultiDrizzle}}
\newcommand\hyperz{\texttt{HyperZ}}

\begin{document}
\title{A Multiply Imaged Luminous Infrared Galaxy Behind the Bullet Cluster
(1E0657-56)\footnotemark[1]} \footnotetext[1]{This paper includes data
gathered with the 6.5 meter Magellan Telescopes located at Las Campanas
Observatory, Chile, the Hubble Space Telescope, and the Spitzer Space
Telescope.}
  
\author{Anthony H. Gonzalez}
\affil{ Department of Astronomy, University of Florida, Gainesville, FL 32611-2055}
\email{anthony@astro.ufl.edu}

\author{Douglas Clowe} 
\affil{ Department of Physics \& Astronomy, Ohio University, Clippinger Labs 251B, Athens, OH 45701}

\author{Maru\v{s}a Brada\v{c}}
\affil{Department of Physics, University of California at Santa
Barbara, Santa Barbara, CA 93106}

\author{Dennis Zaritsky}
\affil{Steward Observatory, University of Arizona, 933 North Cherry
Avenue, Tucson, AZ 85721}

\author{Christine Jones and Maxim Markevitch}
\affil{Harvard-Smithsonian Center for Astrophysics, 60 Garden St.,
Cambridge, MA 02138}

\begin{abstract}

We present evidence for a \spitzer-selected luminous infrared galaxy (LIRG)
behind the Bullet Cluster (1E0657-56). The galaxy, originally identified as a
multiply imaged source using \irac~photometry, has a spectral energy
distribution consistent with a highly extincted ($A_V\sim3.3$), strongly
star-forming galaxy at $z=2.7$.  Using our strong gravitational lensing model
presented in \citet{bradac2006}, we find that the magnifications are
$|\mu|\approx10-50$ for the three images of the galaxy. The brightest and
faintest images differ by a factor of 3.2 in magnification. The implied
infrared luminosity is consistent with the galaxy being a LIRG, with a stellar
mass of $M_*\sim2\times10^{10}$ M$_\odot$ and a star formation rate of
$\sim90$ M$_\odot$ yr$^{-1}$.  With lensed fluxes at 24$\mu$m of 0.58 mJy and
0.39 mJy in the two brightest images, this galaxy presents a unique
opportunity for detailed study of an obscured starburst with a star formation
rate comparable to that of $L^*$ galaxies at $z>2$.

\end{abstract}


\begin{deluxetable*}{lllllllll}
\tabletypesize{\scriptsize}
\tablecaption{Observed Fluxes and Magnitudes for Lensed Images}
\tablehead{
\colhead{}       &\multicolumn{2}{c}{Image A}&\multicolumn{2}{c}{Image B} &\multicolumn{2}{c}{Image C} & \multicolumn{2}{c}{Flux Ratios}   \\
\colhead{Passband}       &\colhead{Flux ($\mu$Jy)}  &\colhead{Mag (AB)}     &\colhead{Flux($\mu$Jy)}        & \colhead{Mag (AB)} &\colhead{Flux($\mu$Jy)}        & \colhead{Mag (AB)}  & $B/A$ & $B/C$
}
\startdata
F606W\tablenotemark{a} & $<$0.29 & $>$25.26 & $<$0.29 & $>$25.26&$<$0.29 & $>$25.26&\nodata&\nodata\\
F775W & $<$0.24 & $>$24.67 & $<$0.24 & $>$24.67&$<$0.24 & $>$24.67 &\nodata&\nodata\\
F850LP & $<$0.41 & $>$24.11 & $<$0.41 & $>$24.11&$<$0.41 & $>$24.11 &\nodata&\nodata\\
J$_c$ & $<$1.30 & $>$23.62 & $<$1.30 & $>$23.62&$<$1.30 & $>$23.62 &\nodata&\nodata\\
K$_s$ & $<$5.35 & $>$22.08 & $<$5.35 & $>$22.08&$<$5.35 & $>$22.08 &\nodata&\nodata\\
3.6$\mu$m\tablenotemark{b} & $13.6\pm0.6$ & $21.06\pm0.05$& $21.0\pm1.5$& $20.59\pm0.08$ & $7.4\pm1.7$ & $21.73\pm0.25$ & $1.54\pm0.13$ & $2.84\pm0.68$\\
4.5$\mu$m & $23.4\pm0.9$ & $20.47\pm0.04$ & $32.7\pm1.5$& $20.11\pm0.05$ & $10.6\pm1.7$ & $21.34\pm0.18$ & $1.40\pm0.08$ & $3.08\pm0.52$\\
5.8$\mu$m & $38.4\pm1.7$ & $19.94\pm0.05$ & $57.9\pm1.9$& $19.49\pm0.03$ & $16.8\pm1.7$ & $20.84\pm0.11$ & $1.51\pm0.08$ & $3.45\pm0.37$ \\
8$\mu$m & $46.4\pm2.5$ & $19.73\pm0.06$ & $67.1\pm2.5$ & $19.33\pm0.04$ & $20.6\pm2.5$ & $20.62\pm0.13$ & $1.45\pm0.09$ & $3.25\pm0.41$\\
24$\mu$m & $390\pm20$ & $17.42\pm0.05$  &$575\pm20$ & $17.00\pm.04$ & $175\pm20$ & $18.29\pm0.12$ &$1.47\pm0.09$ &$3.29\pm0.39$\\
\enddata
\tablenotetext{a}{All quoted upper limits are 5-$\sigma$ confidence.}
\tablenotetext{b}{All \irac~photometry is calculated within 2.4$\arcsec$ 
apertures and corrected to total magnitudes using the published point source aperture corrections. No additional correction has been applied to account for the extended nature of the source.}
\label{tab:photometry}
\end{deluxetable*}

\section{Introduction}
\label{sec:intro}

Measurements of both the star formation history of the universe and corollary
build-up of stellar mass have established that the star formation rate peaks
at $1\la z\la 3$
\citep[e.g.][]{madau1996,lilly1996,dickinson2003,rudnick2003,reddy2008,wilkins2008}.
\spitzer~24$\mu$m observations further indicate that star formation at this
epoch is dominated by luminous and ultraluminous infrared galaxies
\citep[LIRGs and ULIRGs;][]{perez2005,lefloch2005}.  ULIRGs are sufficiently
bright to facilitate spectroscopy and detailed analyses
\citep{daddi2005,yan2005,valiante2007,pope2008}; however, these galaxies
represent only the most massive tail of the galaxy population
\citep{dey2008,dye2008}.

In contrast, LIRGs have properties more similar to the overall galaxy
population, with stellar masses and star formation rates comparable to those
seen for UV-selected star-forming galaxies at $z\sim2$
\citep{reddy2004,reddy2006}.  The intrinsic faintness of LIRGS however
precludes both high-fidelity multiwavelength photometry and spectroscopic
programs at optical and infrared wavelengths at $z\ga2$.

Strong gravitational lensing enables observations of intrinsically fainter
galaxies than is otherwise possible, and several recent programs have begun to
exploit lensing by galaxy clusters to probe the properties of infrared and
submillimeter luminous galaxies \citep[e.g.][ and references
therein]{Knudsen2008, rigby2008}. The main limitation of this approach is
simply the small number of known lensed galaxies that are luminous at these
wavelengths.

In this paper we present evidence for a strongly lensed, luminous infrared
galaxy that is triply imaged by the Bullet Cluster. This galaxy is the only
strongly lensed source for which the initial detection was made with
\spitzer~at mid-infrared wavelengths, and due to its large magnification
provides a window onto the properties of lower luminosity infrared galaxies
than have previously been studied in this redshift regime.  In previous papers
our team has explored the physical properties of the Bullet Cluster,
1E0657-56, measuring its matter distribution and the properties of the X-ray
gas \citep{markevitch2002,markevitch2004,clowe2004,clowe2006,bradac2006}, and
constraining the dark matter self-interaction cross-section
\citep{markevitch2004,randall2007}.  The object that is the subject of this
paper was first identified as a doubly lensed source in one of these papers
\citep[][\S 6 and Fig. 8]{bradac2006}, and is independently detected as a
millimeter source \citep{wilson2008a}.  Here we incorporate new \hst,
\spitzer, and Magellan data, in a detailed analysis of this object.  The data
are presented in \S \ref{sec:data}, and are used in \S \ref{sec:analysis} to
estimate the redshift, magnification, stellar mass, and star formation rate.
In this section we also present evidence for a newly discovered, third image
of this galaxy.  We summarize our results in \S \ref{sec:conclusions}.

\begin{figure*}
\begin{center}
\epsscale{0.99} \plotone{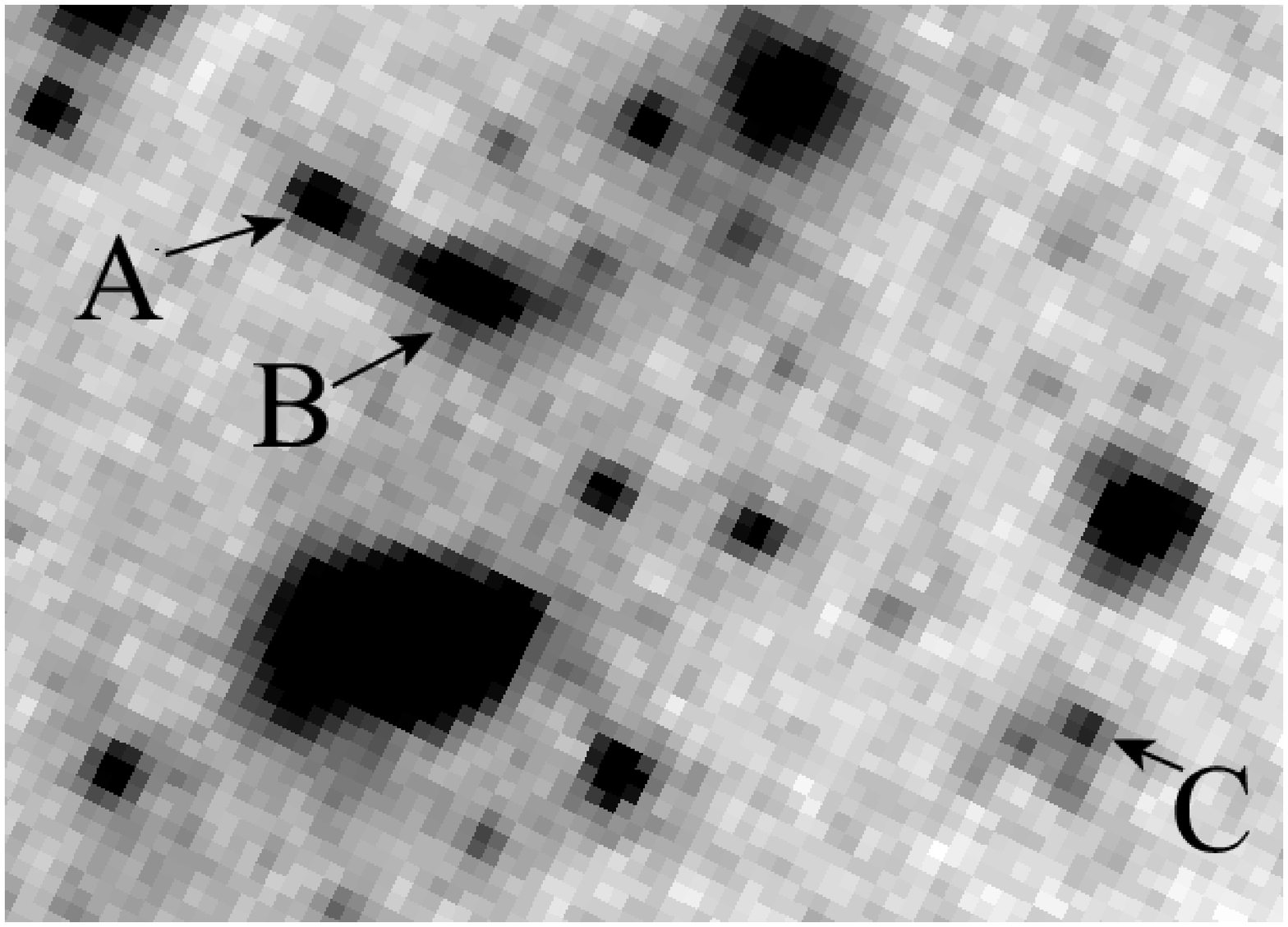}
\epsscale{0.49} \plotone{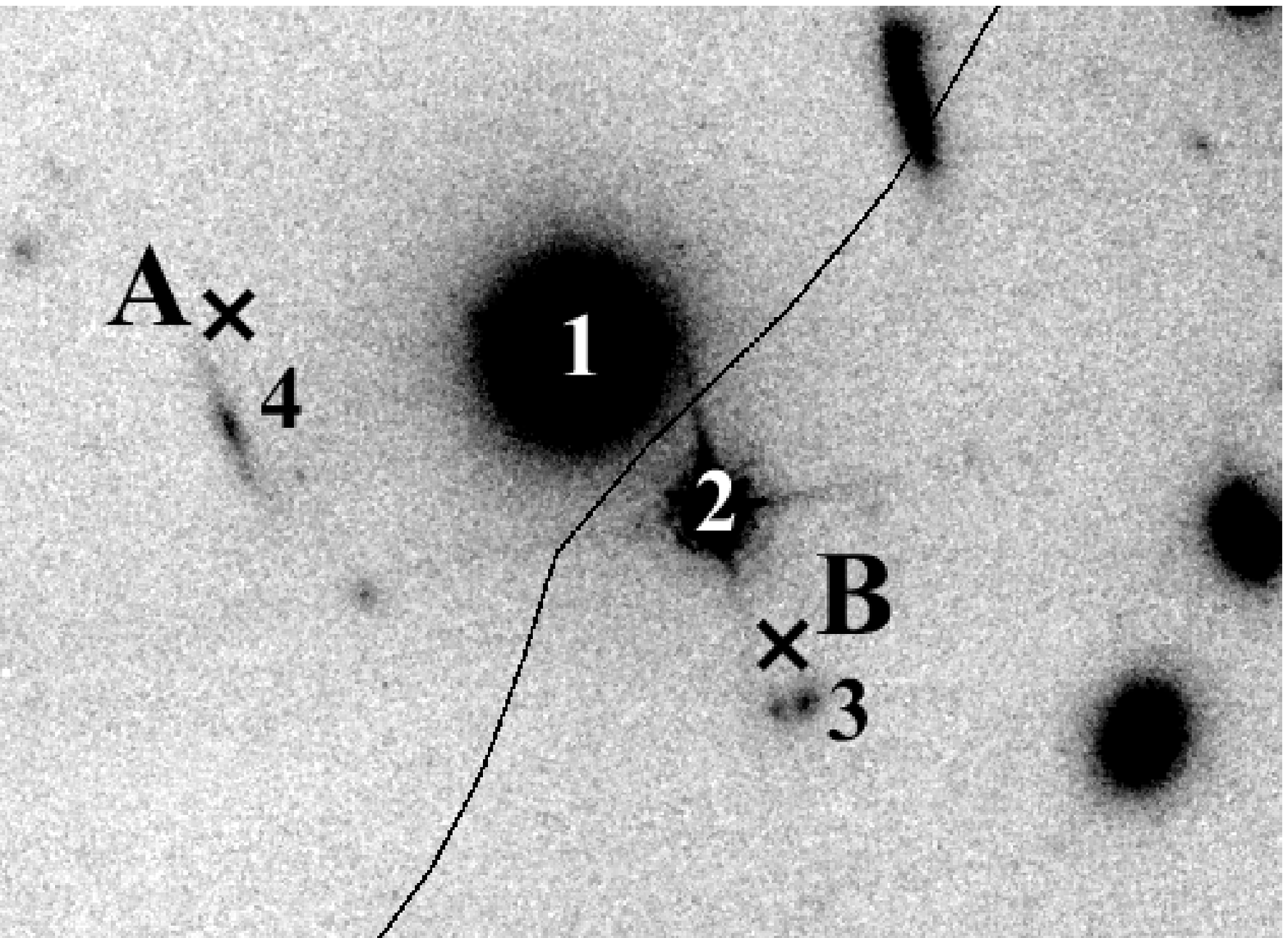}\plotone{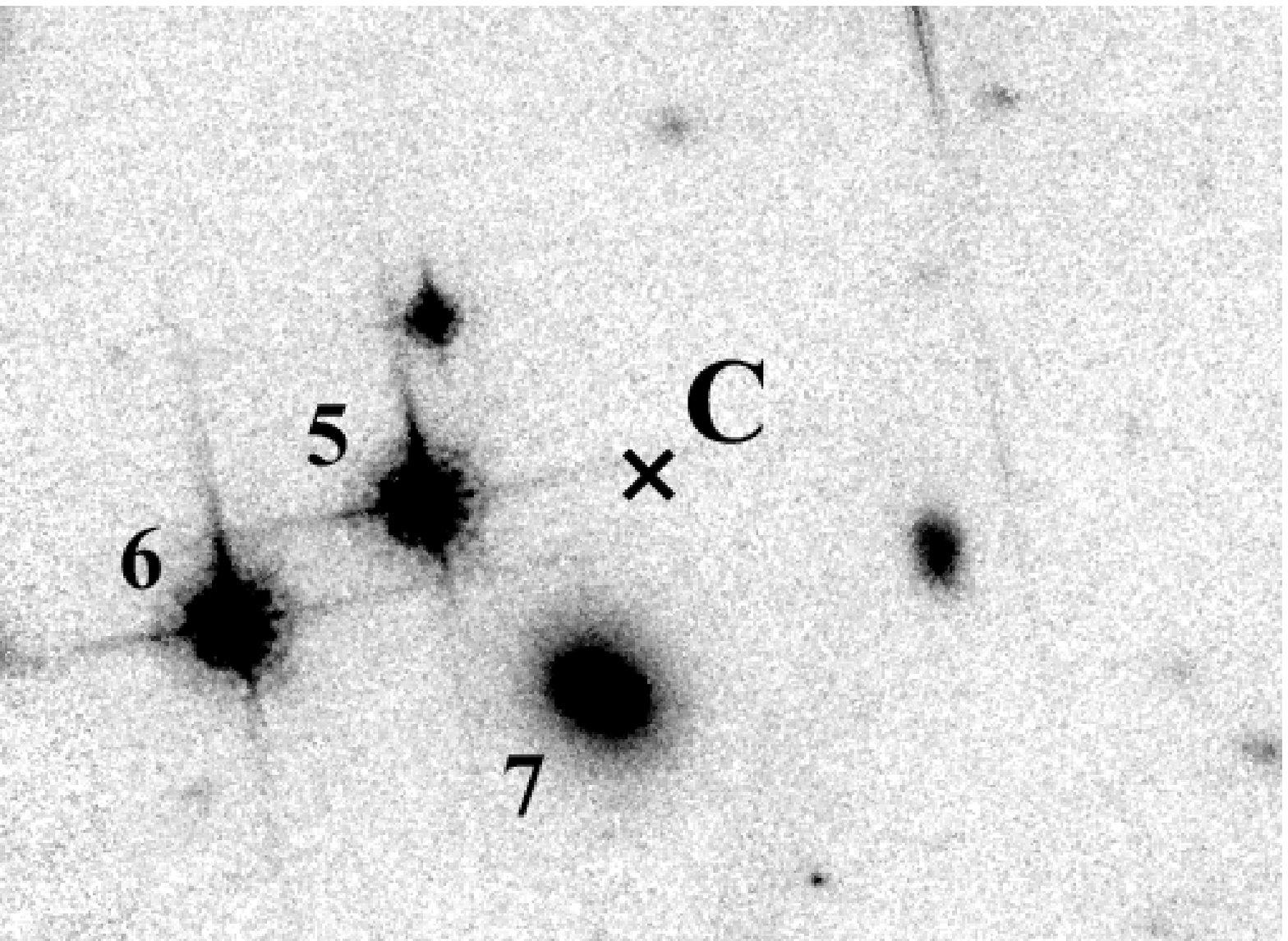}
\epsscale{1.00} 
\end{center}
\caption{{\it Top--} An 8$\mu$m image showing the location of image C relative
to images A and B. In this image the galaxy between images A and B has been
subtracted for clarity. The field of view is $65\arcsec\times50\arcsec$.  {\it
Bottom --} F850LP cutouts of the regions surrounding lensed images A and B
(left) and image C (right). The crosses denote the locations of each image;
the solid curve in the left panel is the critical curve from the $z=2.7$
magnification map. The objects detected in the F850LP image that lie closest
to the \irac~coordinates for images A and B are numbered 1-4. Object 1 is a
cluster elliptical galaxy and object 2 is a star. The two fainter galaxies (3
and 4) are offset from the \irac~detections by 0.8$\arcsec$ and 1.5$\arcsec$,
whereas the relative astrometry is good to 0.25$\arcsec$, and can thus be
excluded as optical counterparts to the lenses. The field of view is
$17\arcsec\times12\arcsec$ in both panels. For all images north is up and east
is to the left.\label{fig:bigimg}}
\end{figure*}

\section{Photometric Data and Measurements}
\label{sec:data}

We use the combination of \spitzer~\irac~\citep{Fazio2004} and
\mips~\citep{Rieke2004}, \hst~\acs~\citep{Ford2003}, and ground-based
near-infrared (NIR) observations to constrain the spectral energy distribution
of the multiply imaged source. The subsections below describe the data and
photometric analysis associated with each facility.

\subsection{\spitzer~\irac}

We originally detected this source as a doubly imaged object in
\spitzer~\irac~data obtained on December 17-18, 2004. The data from this
program include imaging in all four \irac~bands (3.6$\mu$m, 4.5$\mu$m,
5.8$\mu$m, and 8$\mu$m). These data were taken with a cycling dither pattern
with medium scale factor and 100s frame time during an 8720s duration
Astronomical Observation Request (AOR).  The effective exposure times are 4 ks
in each filter.

We process the data using MOPEX \citep{Makovoz2006}, with a final pixel scale
of $0\farcs 86$.  Before measuring aperture fluxes, we first use GALFIT
\citep{peng2002} to model and subtract a cluster elliptical that lies directly
between the two lensed images (object \#1 in Figure \ref{fig:bigimg}), using a
nearby, isolated star as the input PSF for GALFIT. The structural parameters
for the galaxy are held fixed to values derived using the \acs~data (see
below), with only the position and magnitude permitted to vary.\footnote{The
position is permitted to vary at the subpixel level to minimize errors in the
subtraction due to residual mis-registratioin between the images. Fixing the
position does not qualitatively alter our results.}  There is also a star
between the two lensed images (object \#2 in Fig. \ref{fig:bigimg}) that is
detected at 3.6$\mu$m and 4.5$\mu$m, but is fainter at these wavelengths than
the lensed images. Due to its faintness, we mask this star rather than model
it with GALFIT.  We also mask two other nearby sources that lie within the
background apertures (\#3 and \#4 in Fig. \ref{fig:bigimg}).

\begin{figure*}
\epsscale{0.24}
\plotone{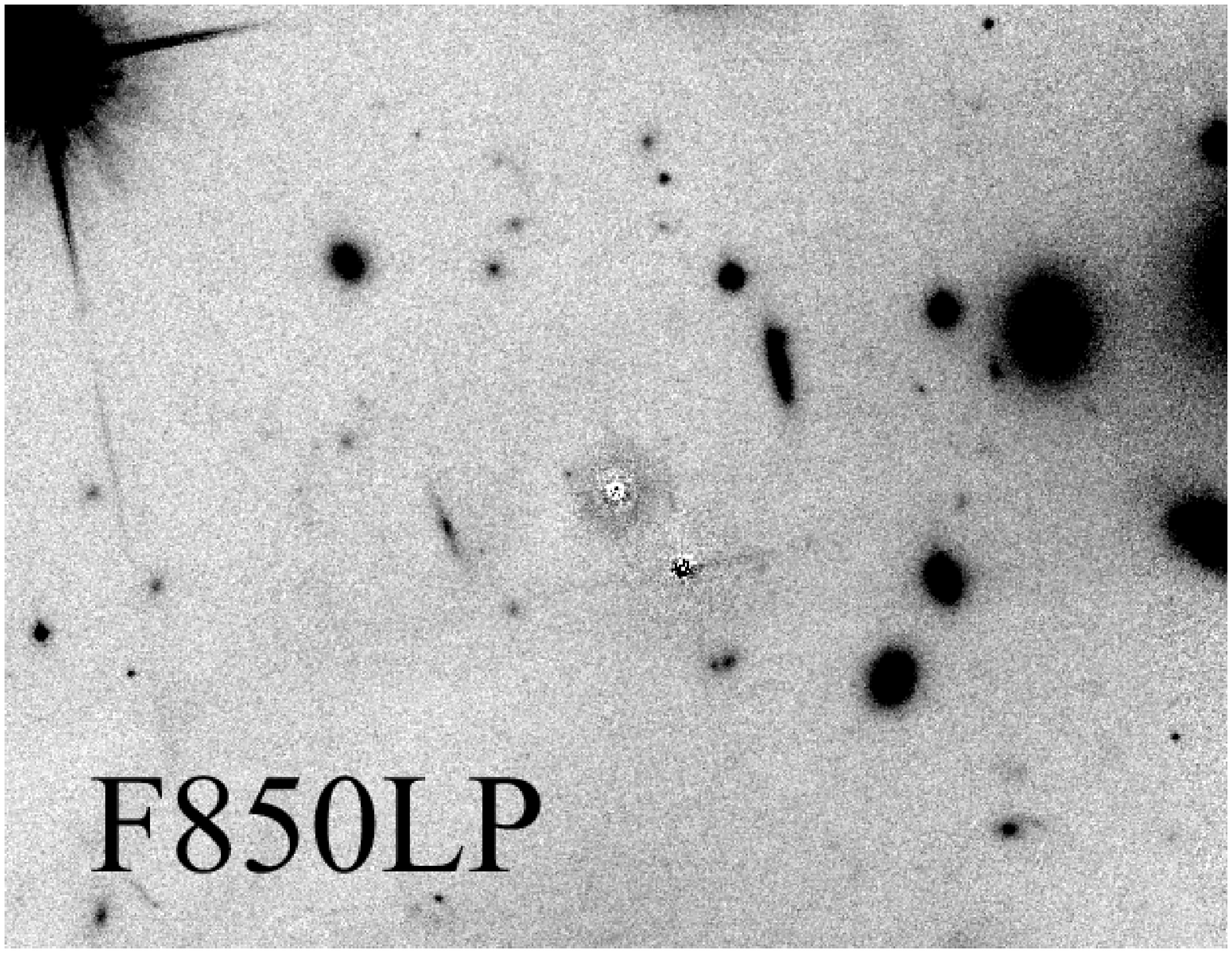}\plotone{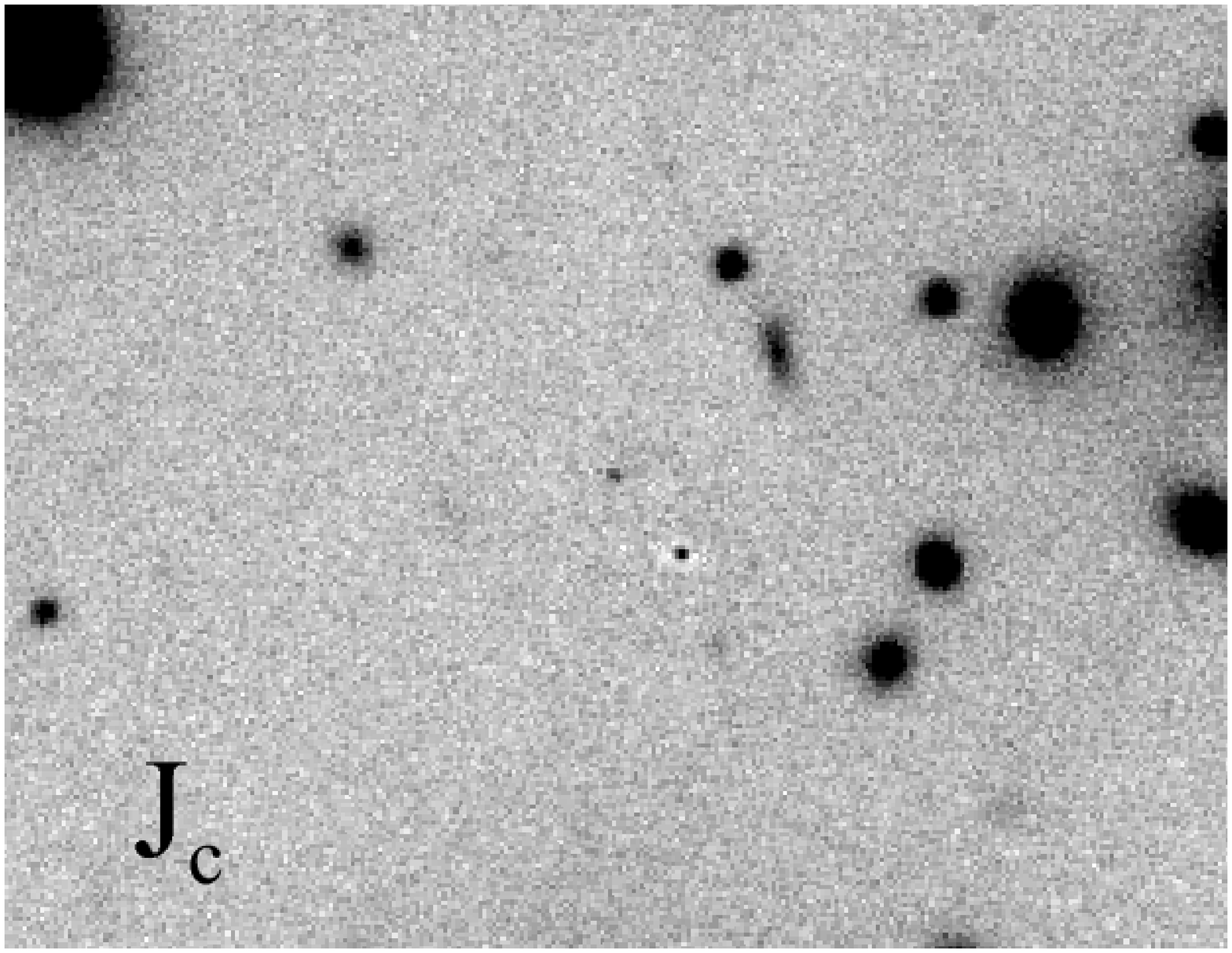}\plotone{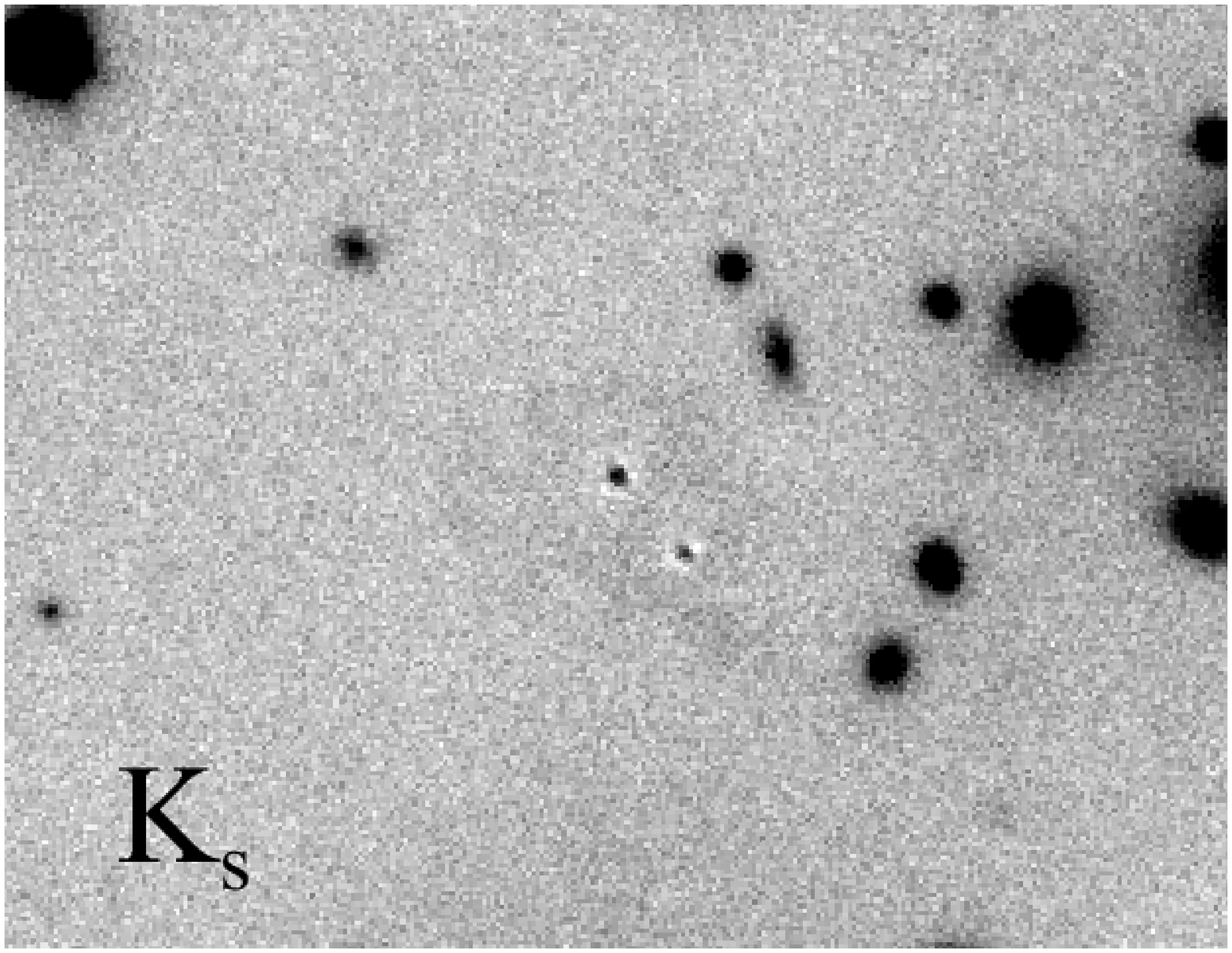}\plotone{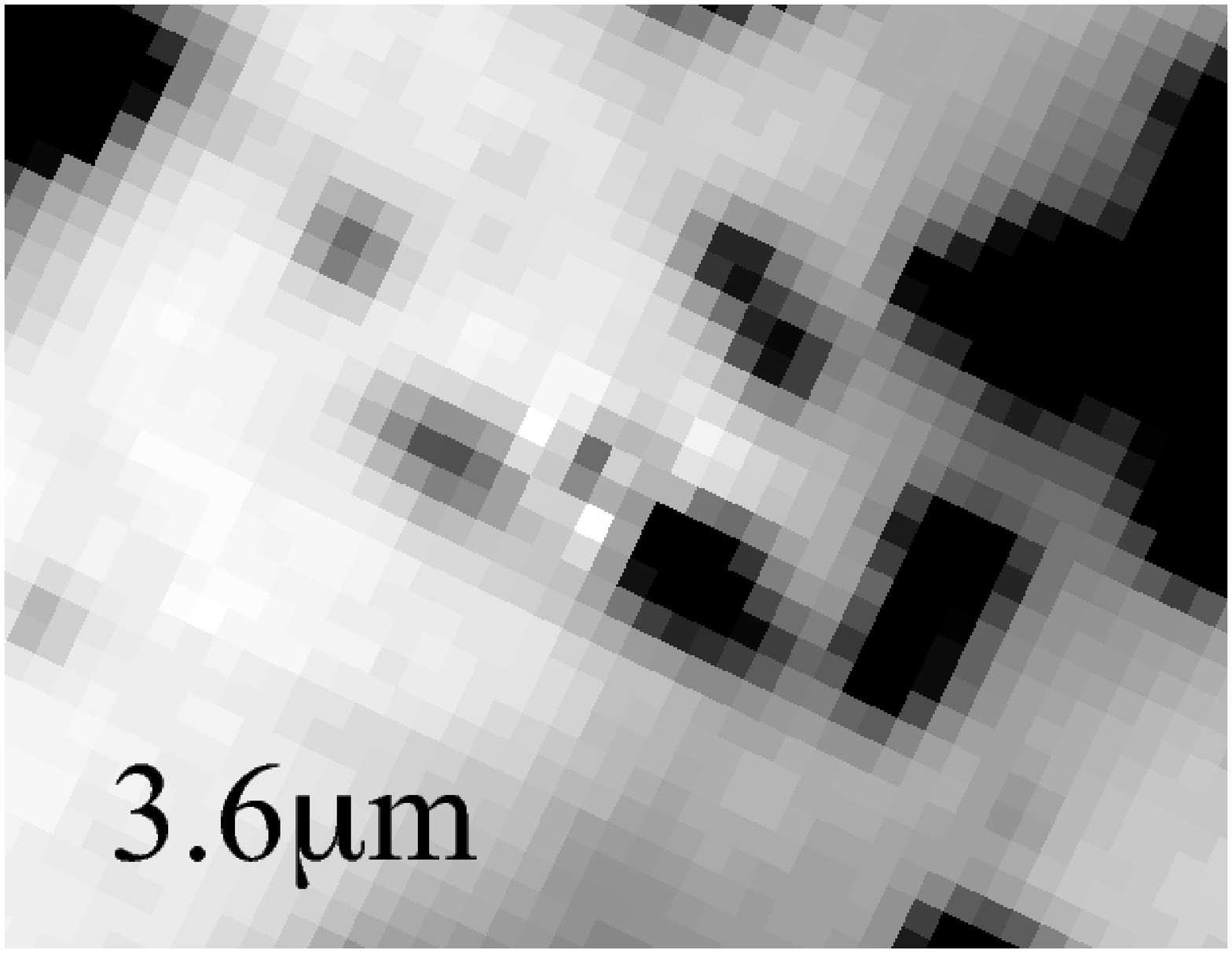}\\
\plotone{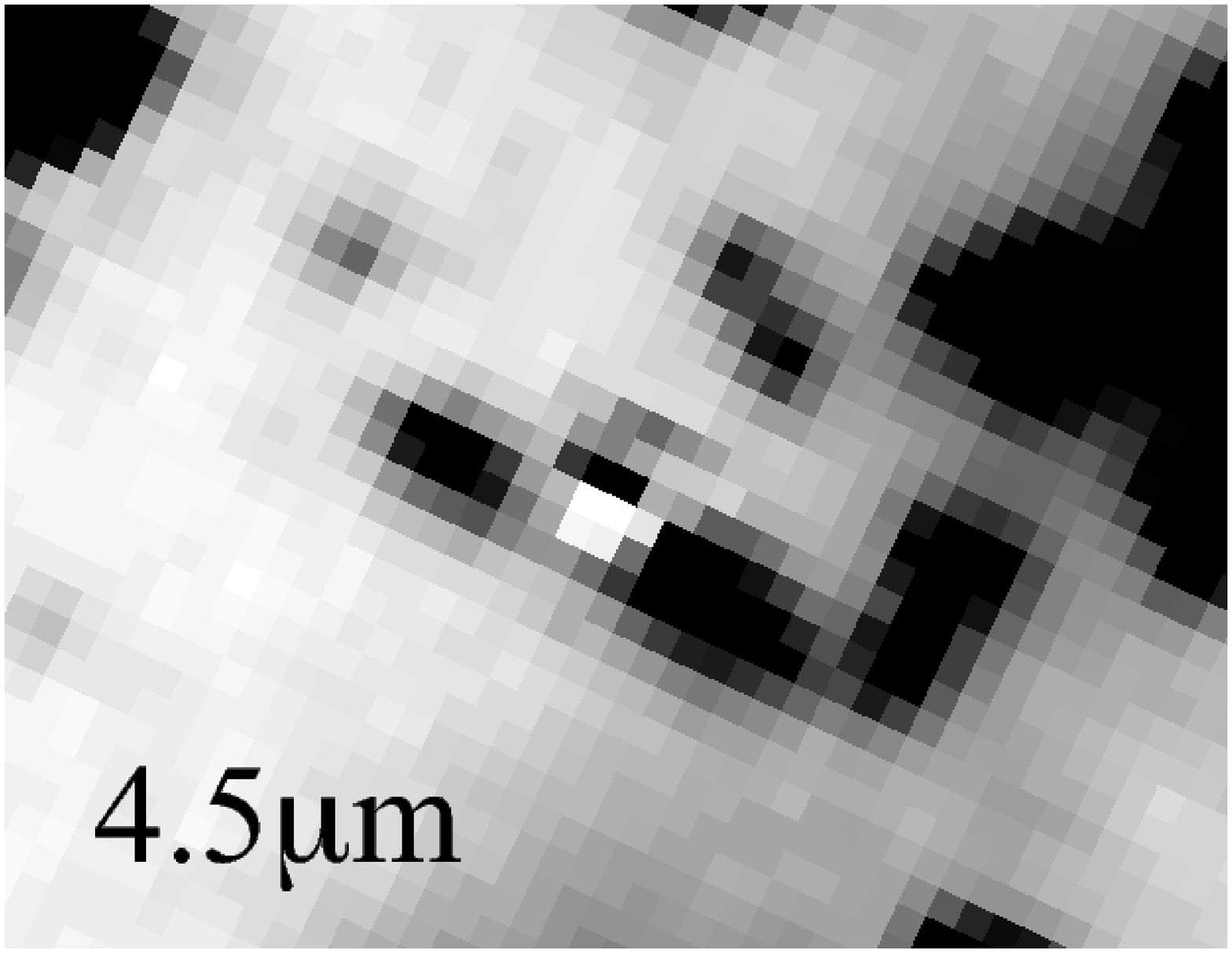}\plotone{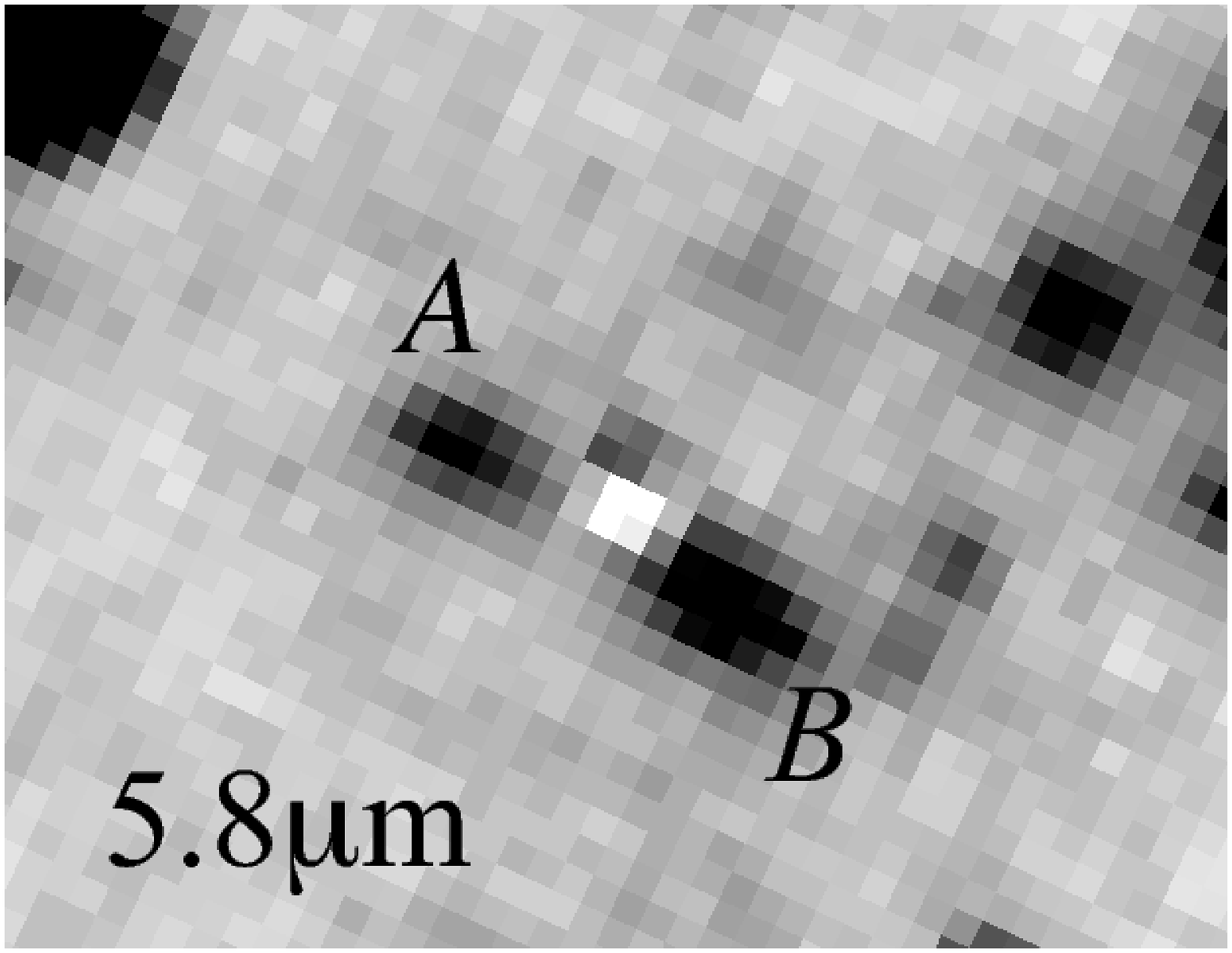}\plotone{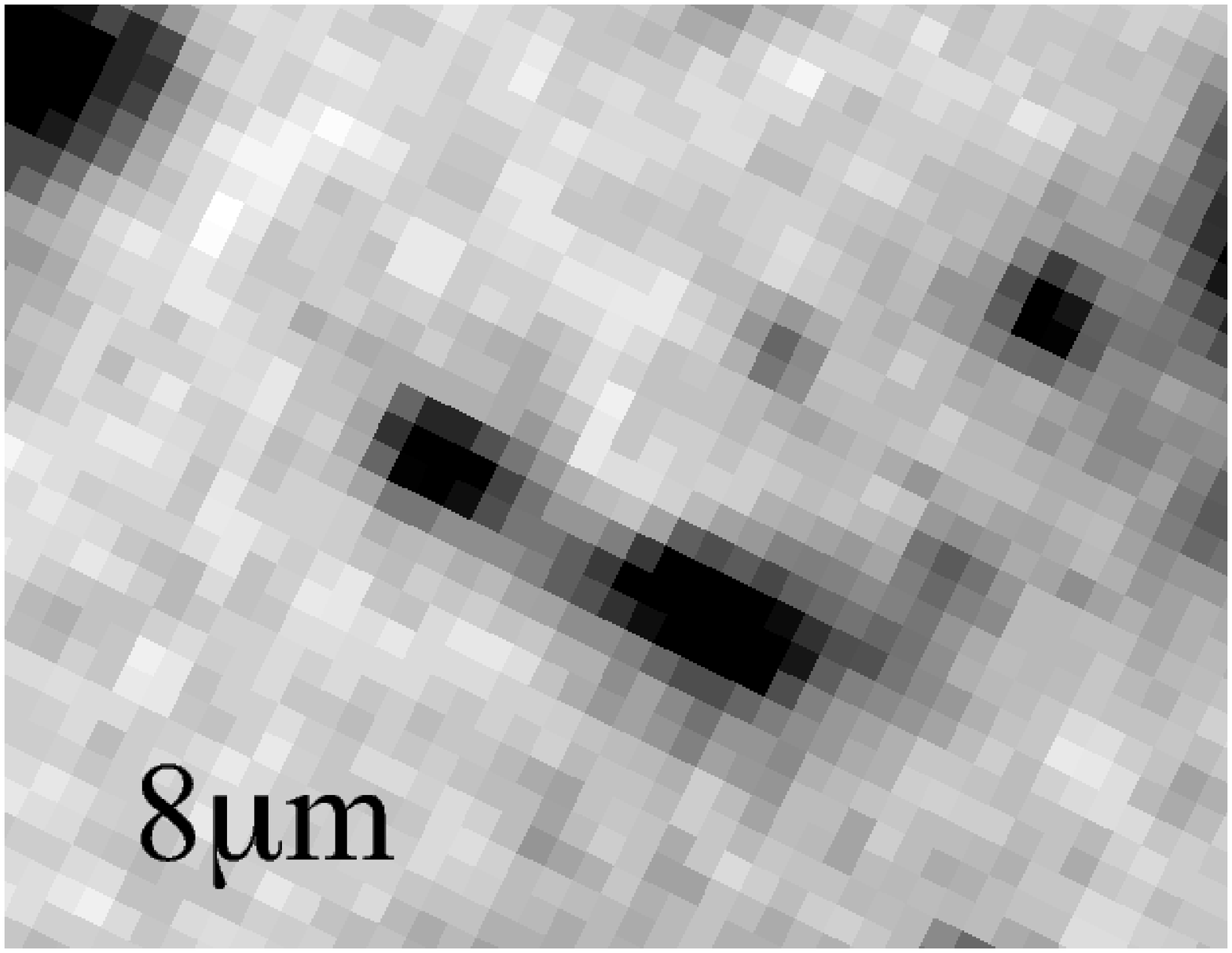}\plotone{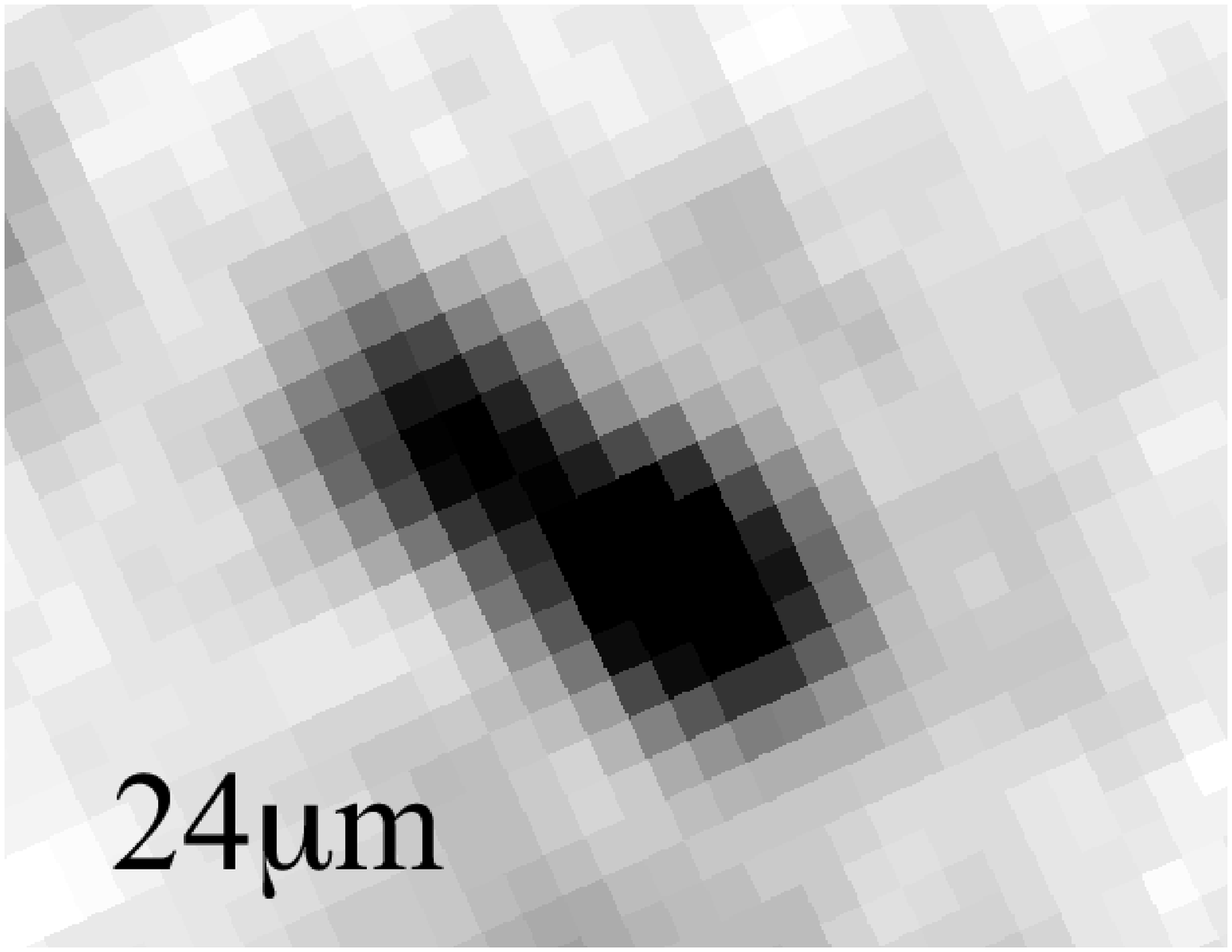}\\
\vskip 0.3cm
\plotone{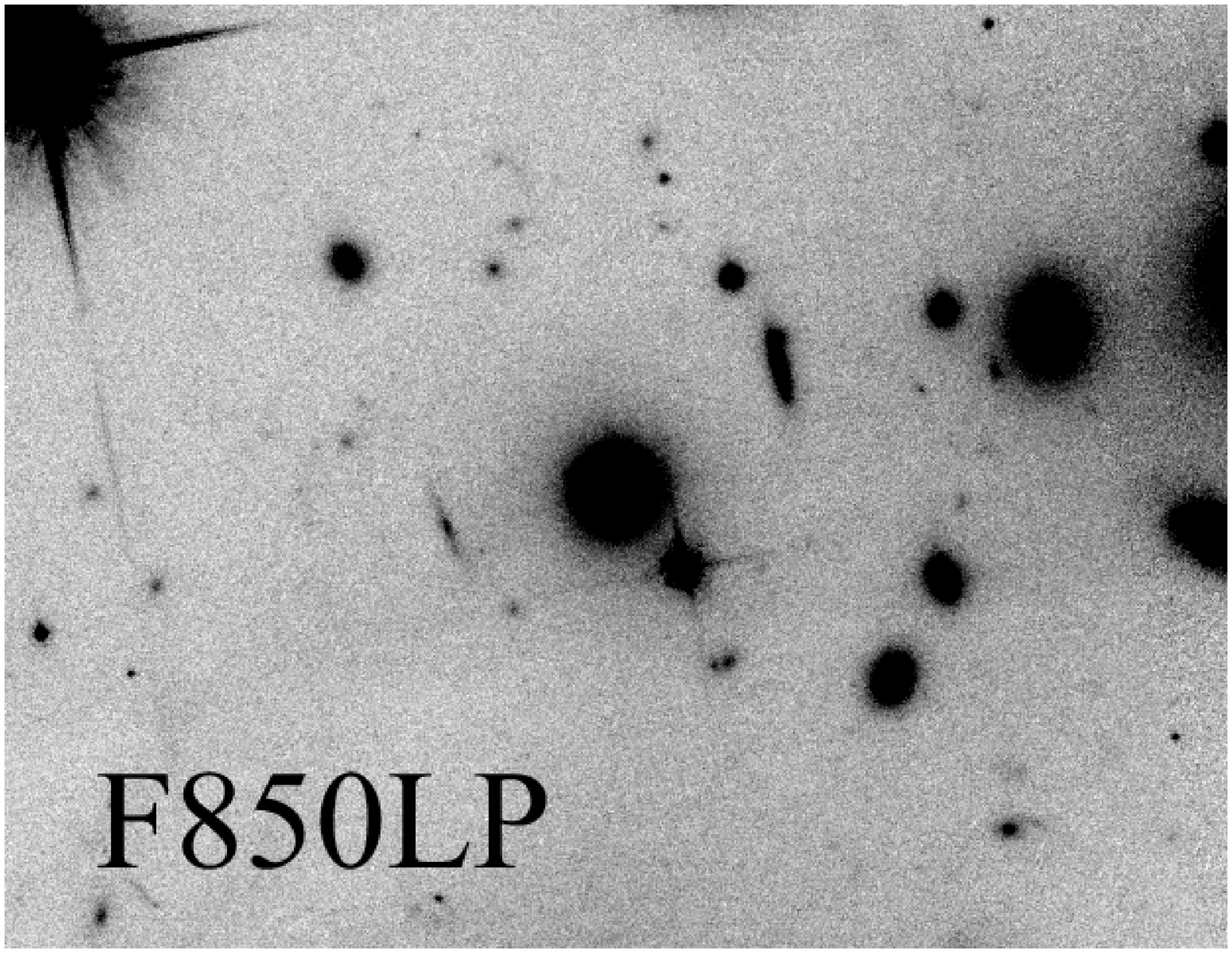}\plotone{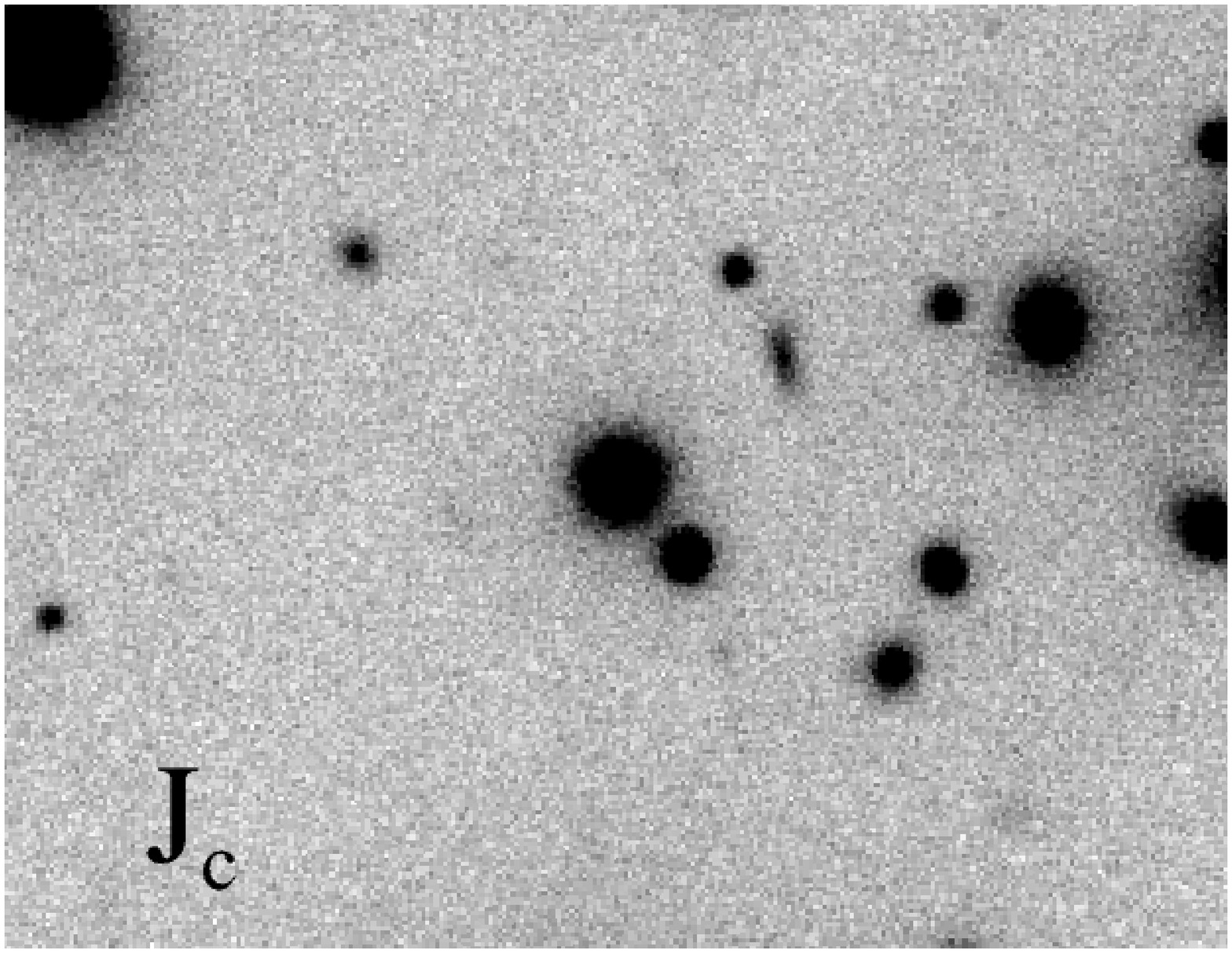}\plotone{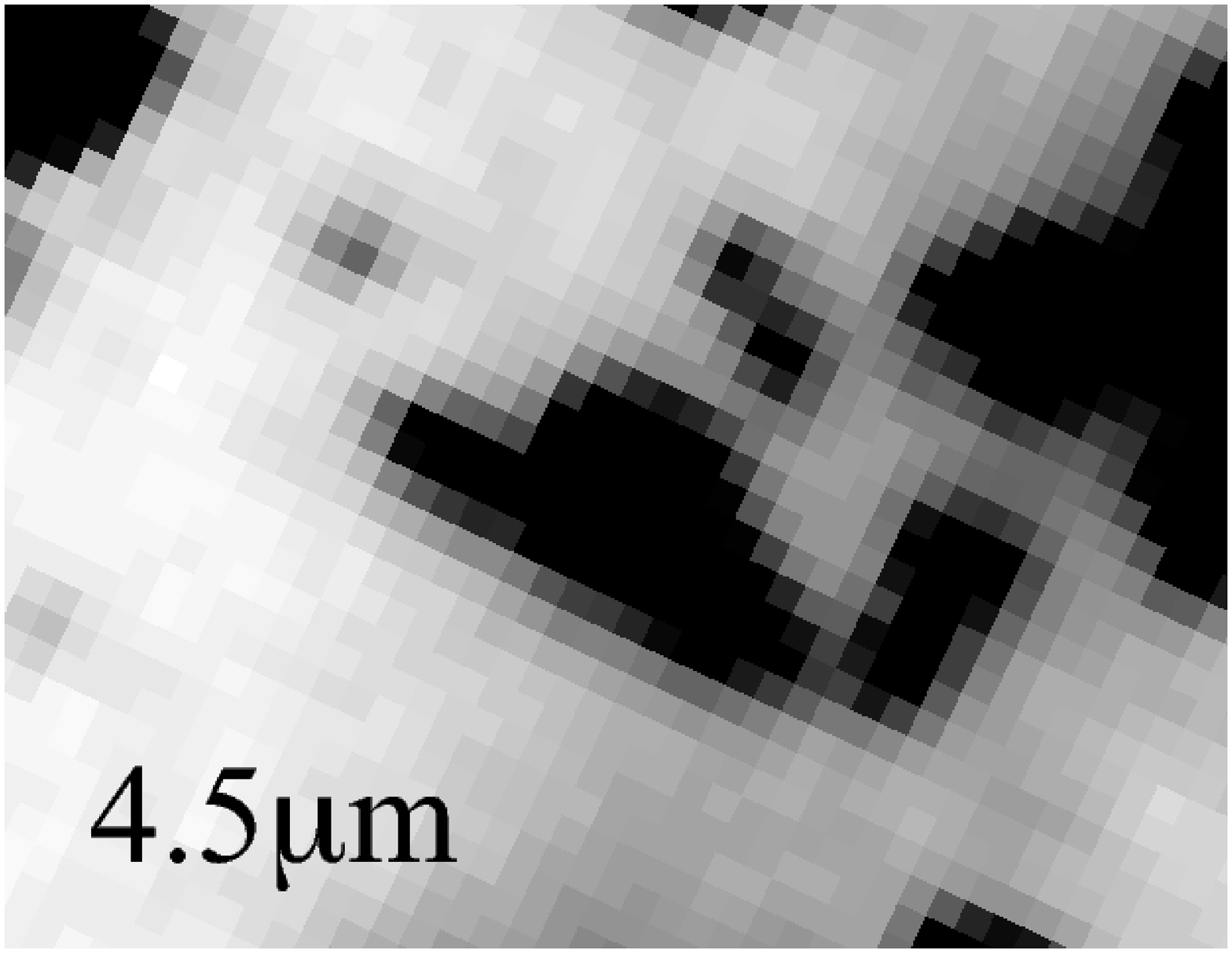}\plotone{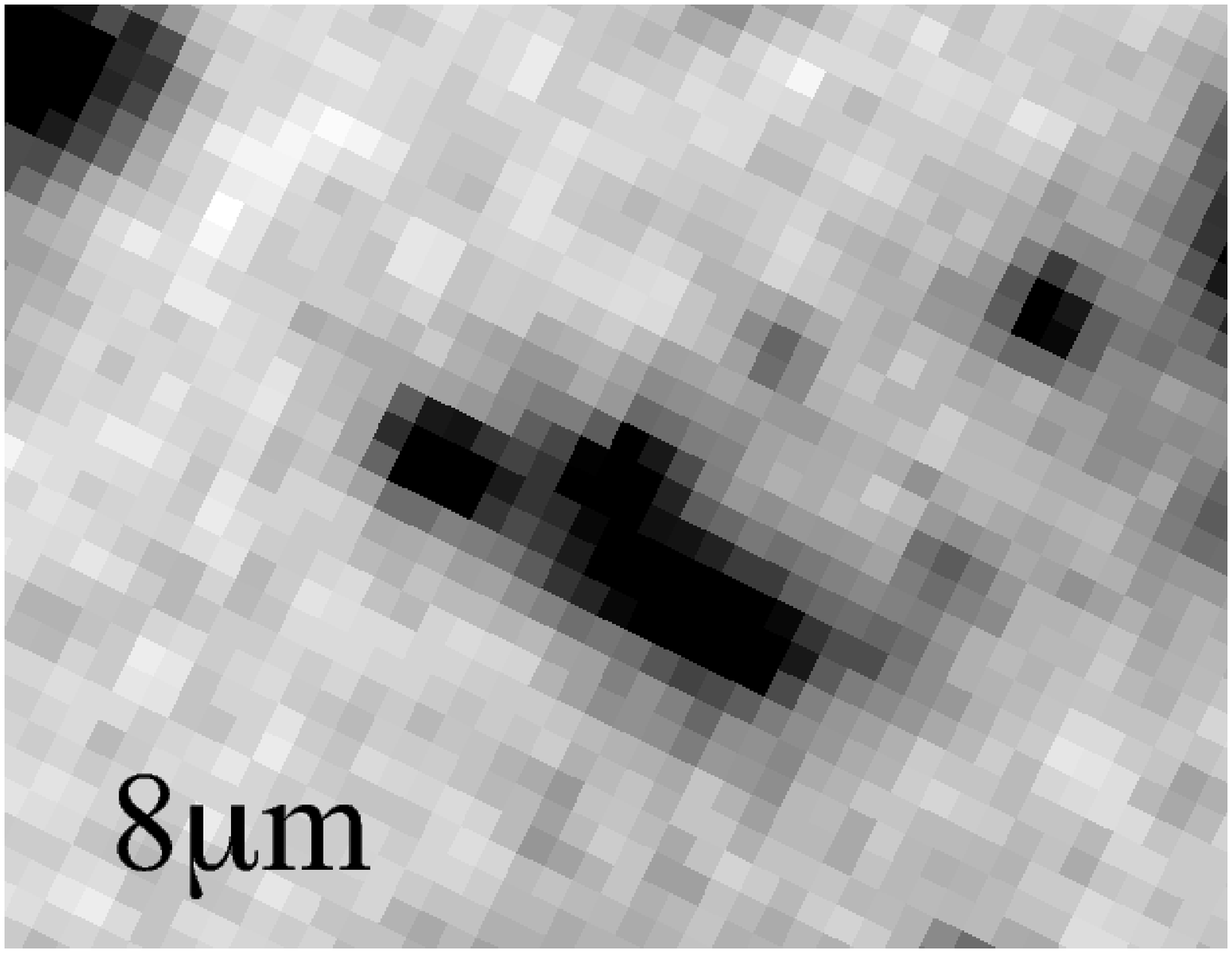}\\
\caption{Cutout images of the region around the lensed galaxy, in order of
increasing wavelength starting with the ACS F850LP data. In the first two rows
we show the images used for the photometry after subtraction of contaminant
sources using GALFIT.  In the last row we show images prior to GALFIT
subtraction for a representative sample of wavelengths
(F850LP,$J$,4.5$\mu$m, and 8$\mu$m).  In all panels the field of view is
$34\arcsec\times26\arcsec$. North is up and east is to the
left. \label{fig:images}}
\end{figure*}

\begin{figure}
\epsscale{1}
\plotone{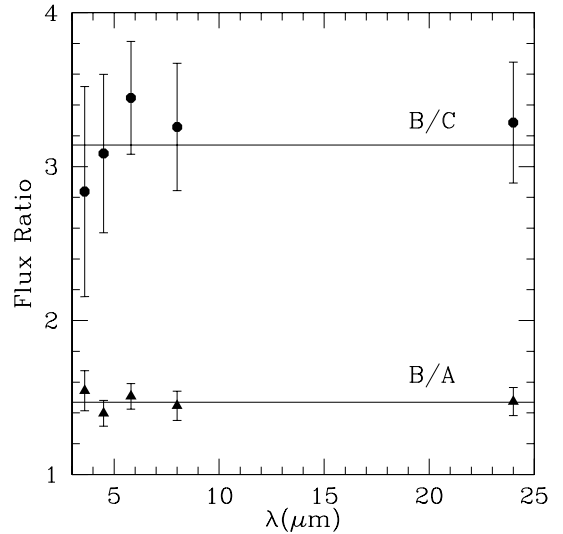}
\caption{Flux ratios for the three images as a function of wavelength.  The
solid lines correspond to the weighted mean flux ratios, including data at all
wavelengths.  The observed flux ratios at different wavelengths are consistent
to within the photometric uncertainties For B/A (triangles) and B/C (circles)
these mean values are $1.47\pm0.06$ and $3.18\pm0.23$ respectively. The flux
ratio for A/C, which is not plotted, is $2.16\pm0.18$.
\label{fig:fluxratio}}
\end{figure}

We then perform aperture photometry using an aperture of radius 2.4$\arcsec$,
with a background annulus extending from $2.4-7.2\arcsec$, applying the point
source aperture corrections given in the \irac~Data Handbook.\footnote{See
\url{http://ssc.spitzer.caltech.edu/irac/dh/}. A 2.4$\arcsec$ aperture is the
smallest for which corrections are given.}  We note that the total magnitude
may be underestimated by $\sim10$\% due to the modest spatial extension of the
source in the \irac~data, but apply no additional correction for this factor.
We measure the flux in an ensemble of off-source apertures to compute the
photometric uncertainties. The resultant photometry is given in Table
\ref{tab:photometry} for these two images, which we denote as A and B
(Fig. \ref{fig:bigimg} and \ref{fig:images}). Image B is brighter than A by a
factor of 1.5, with consistent flux ratios in all bands
(Fig. \ref{fig:fluxratio}), as required for a multiply imaged source. The
coordinates are $(\alpha_{2000},\delta_{2000})=$(06:58:38.0,-55:57:02) for
image A and $(\alpha_{2000},\delta_{2000})$=(06:58:37.1,-55:57:06) for image
B.

In Table \ref{tab:photometry} we also present photometry for a newly
discovered third image of the galaxy, hereafter denoted as image C (see Fig
\ref{fig:bigimg}), which is discussed in greater detail in \S
\ref{sec:magnification}. The \irac~photometry for image C is obtained in the
same fashion as for images A and B, in this case modelling and subtracting two
nearby stars and one nearby galaxy (objects \#5-7 in Fig. 1).

\subsection{\spitzer~\mips}
\label{sec:mipsdata}

The \mips~24$\mu$m data were acquired on November 30, 2007. Observations were
taken in small scale photometry mode using a 3$\times$3 raster map with each
position offset by half the array. The frame time for individual exposures is
30s.

We process the data using MOPEX, with a final pixel scale of $1\farcs 22$. At
this wavelength images A, B, and C are all clearly detected.  We use APEX
\citep[Astronomical Point source EXtractor;][]{Makovoz2002,Makovoz2005} to
perform point source photometry, deblending images A and B. The fluxes of all
three lensed images, which are derived via PSF fitting, are reported in Table
\ref{tab:photometry}. The quoted uncertainties include uncertainty associated
with background subtraction, which dominate over the statistical uncertainties
reported by APEX.  The \mips~point-source photometry does not suffer from
contamination by foreground sources. The spectral energy distributions of
cluster ellipticals fall rapidly between 8$\mu$m and 24$\mu$m and the nearby
sources modelled at shorter wavelengths are not detected at 24$\mu$m.

\subsection{\hst~\acs}

  We observed the Bullet Cluster with the Advanced Camera for Surveys (ACS)
through the F606W, F775W, and F850LP filters.  The F606W data were taken on
October 21, 2004, with a total exposure time of 4.7ks. The F775W and F850LP
data were taken on October 12-13, 2006, with total exposure times of 10.1ks
and 12.7ks, respectively. All observations were reprocessed and drizzled to a
common coordinate system using custom software (Haggles; Marshall et al. 2008,
in prep) based upon \multidrizzle~\citep{koekemoer2002}.

For the \hst~data we perform aperture photometry within 1.5$\arcsec$ aperture
radii. This aperture size is selected as a tradeoff between two competing
factors.  Specifically, while smaller apertures yield more stringent lower
limits on the magnitude, the aperture size must be sufficiently large to
encompass the total flux from each lensed image. Given that \irac~provides the
highest resolution data in which the source is detected, our information about
the true physical extent of the lensed images is limited. The galaxy shows
only limited spatial extent at 3.6$\mu$m, indicating that a 1.5$\arcsec$
aperture is sufficiently large to enclose the total flux in the
\acs~imaging. Any future higher resolution detection of this object will
enable use of smaller apertures and yield improved magnitude limits.  As with
the \irac~data, we determine the photometric uncertainty using an ensemble of
background apertures.

Before measuring aperture fluxes, we again use GALFIT to model and subtract
contaminant sources near images A and B.  In the \hst~imaging this includes
objects \#1 and \#2. Near image C there are no objects that require
subtraction for the \hst~data.  We again use a nearby, isolated star as the
input PSF for GALFIT, which has the advantage over
TinyTim{\footnote{\url{http://www.stsci.edu/software/tinytim/tinytim.html}}}
of correctly reproducing the red halo in F850LP \citep{Gilliland2002}, and we
recover consistent structural parameters for the cluster galaxy in all filters
(effective radius $r_e=0.9\arcsec$ and Sersic index $n=4.6$). We also mask out
galaxy \#3 in the aperture of image B (Fig. \ref{fig:bigimg}).  This galaxy
cannot be the optical counterpart to the \irac~ detection since it offset from
the \irac~detection by 0.8$\arcsec$, whereas the relative astrometry is good
to 0.25$\arcsec$. Moreover, both the consistent flux ratios in all \irac~bands
for images A and B (see Table \ref{tab:photometry} and \S \ref{sec:analysis})
and the location of the critical curve in the lensing model support the
interpretation that these are multiple images of the same source.  In this
case, the relative flux ratio of the two images should also be preserved in
the \hst~data, and we would detect the counterimage in the other aperture at
high confidence if it were the optical counterpart. For completeness, in Table \ref{tab:nearbyphotometry} we provide the \hst~photometry for the objects that are labelled in Figure \ref{fig:bigimg}, computed using Source Extractor.

\subsection{Magellan PANIC}

We imaged the central region of the Bullet Cluster with the PANIC instrument
\citep{martini2004} on Magellan on March 06, 2006.  Data were obtained in the
$J_c$ and $K_s$ filters and photometrically calibrated to the 2MASS point
source catalog \citep{skrutskie2006}, with seeing of $0.55-0.6\arcsec$ in both
bands. Similar to the approach taken with the other data sets, we use GALFIT
to fit and subtract off the bright galaxy and star that lie between the
locations of images A and B in the \irac~data. We then measure the flux within
the same $1.5\arcsec$ apertures employed for the \acs~analysis, recovering
only upper limits at the positions of all three images.

\section{Analysis}
\label{sec:analysis}
\subsection{Spectral Energy Distribution and Photometric Redshift}
\label{sec:sed}

\begin{figure*}
\epsscale{0.8}
\plotone{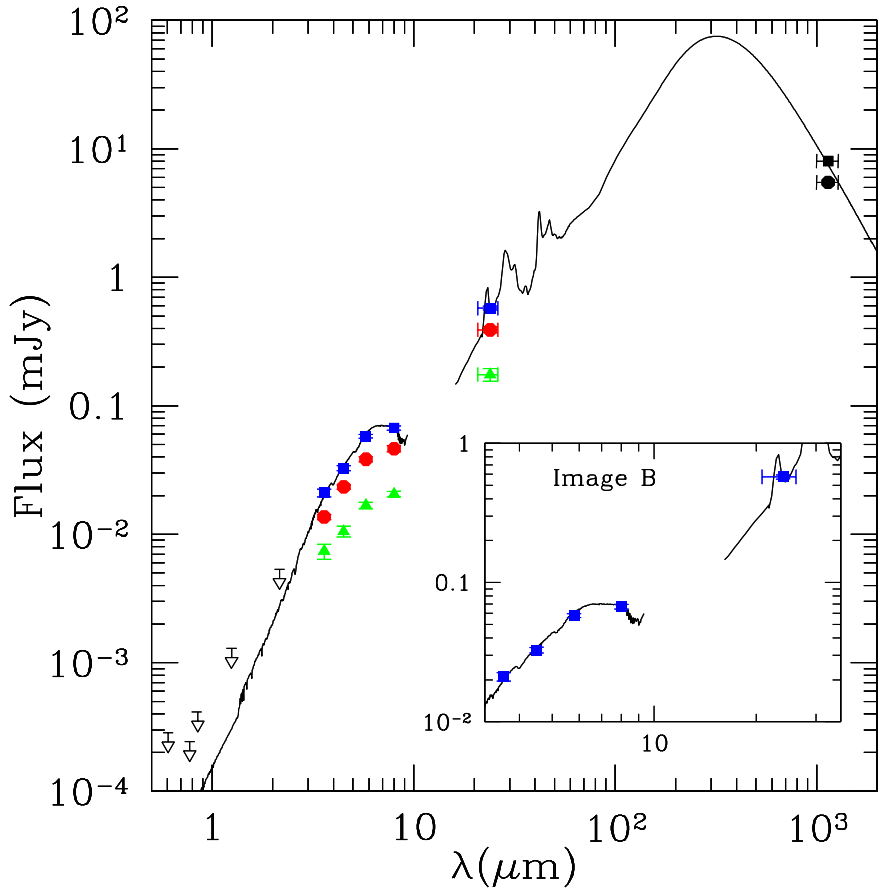}
\caption{Spectral energy distribution for the galaxy. The solid points
correspond to the observed fluxes for images A (circles), B (squares), and C
(triangles), while the open arrows correspond to the 5$\sigma$ upper limits at
optical and NIR wavelengths. The solid circles at 1.1 mm are the AzTEC data
from Wilson et al. (2008), where we have split their observed flux using the
flux ratio derived for images A and B (see Fig. \ref{fig:fluxratio}).
Horizontal error bars on the 24$\mu$m data points denote the width of the
filter, and for the AzTEC data correspond approximately to the system bandpass
\citep{wilson2008b}. The solid curve at $\lambda<10\mu\mathrm{m}$ is the
best-fit spectrum returned by \hyperz~ for image B, which corresponds to a
starburst galaxy.  The solid curve redward of 10$\mu$m is the template from
\citet{chary2001} that best fits the 24$\mu$m flux for image A assuming
$\mu_A=25$. This template is redshifted to $z=2.7$ and rescaled to image B
using the observe flux ratio of the two images
(Fig. \ref{fig:fluxratio}). Note that the AzTEC data, while not included in
the fit, is fully consistent with this model.  The inset zooms in on the
wavelength regime covered by \spitzer, showing only the photometry for image B
for clarity.
\label{fig:sed}}
\end{figure*}

In Figure \ref{fig:sed} we plot the spectral energy distribution (SED) for
each image of the lensed galaxy. Qualitatively the combination of strong upper
limits at optical and near-infrared wavelengths coupled with \irac~detections
and a strong \mips~ detection argue for the galaxy being a dusty starburst at
$z\sim2$, with the \mips~24$\mu$m emission being due to the redshifted PAH
features. The 24$\mu$m emission is difficult to explain if $z\ga3$, while the
galaxy should be detected at NIR or optical wavelengths if either the internal
extinction is low or the redshift is much below 2.
 
\begin{deluxetable}{ccccc}
\tabletypesize{\scriptsize}
\tablecaption{\hst~Photometry for Objects Near Lensed Images}
\tablewidth{0pt}
\tablehead{
\colhead{Fig. 1} & \colhead{F606W} & \colhead{F775W} & \colhead{F850LP} & \colhead{Object}\\
\colhead{ ID} & \colhead{(AB)} & \colhead{(AB) } & \colhead{(AB)} &\colhead{Type}
}
\startdata
1&   $19.89\pm.02$   &  $19.05\pm.03$ & $18.69\pm.03$ & Galaxy\\
2 &  $19.79\pm.01$   &  $18.95\pm.01$ & $19.17\pm.01$ & Star \\
3 &  $24.27\pm.07$   &  $23.43\pm.07$ & $23.39\pm.06$ & Galaxy \\
4 &  $24.42\pm.07$   &  $23.58\pm.08$ & $24.06\pm.08$ & Galaxy \\
5 &  $20.34\pm.01$   &  $19.50\pm.01$ & $18.75\pm.01$ & Star \\
6 &  $19.81\pm.01$   &  $18.97\pm.01$ & $18.60\pm.01$ & Star \\
7 &  $21.27\pm.03$   &  $20.43\pm.03$ & $20.18\pm.03$ & Galaxy\\
\enddata
\tablecomments{In this Table we quote Source Extractor AUTO magnitudes. The
uncertainties are calculated using artificial stars and galaxies.}
\label{tab:nearbyphotometry}
\end{deluxetable}

For a more quantitative answer, we use the photometric redshift code \hyperz~
\citep{bolzonella2000}.  The input spectral templates are obtained using the
Charlot \& Bruzual 2007 models \citep{Bruzual2007} with the Padova 1994
evolutionary tracks \citep{bertelli1994} and a \citet{chabrier2003} mass
function. The templates are defined to have star formation histories identical
to the default synthetic templates provided with \hyperz,\footnote{These
templates have exponentially declining star formation rates with $\tau=1, 2,
3, 5, 15, 30$ Gyr for E, S0, Sa, Sb, Sc, and Sd galaxies, respectively. There
is also a starburst template that corresponds to a single, instantaneous burst
model. For all templates the metallicity is solar; however,
\citet{bolzonella2000} demonstrated that the redshift determination is not
strongly dependent upon metallicity.  } and a \citet{calzetti2000} extinction
law is employed.

\begin{figure}
\epsscale{1}
\plotone{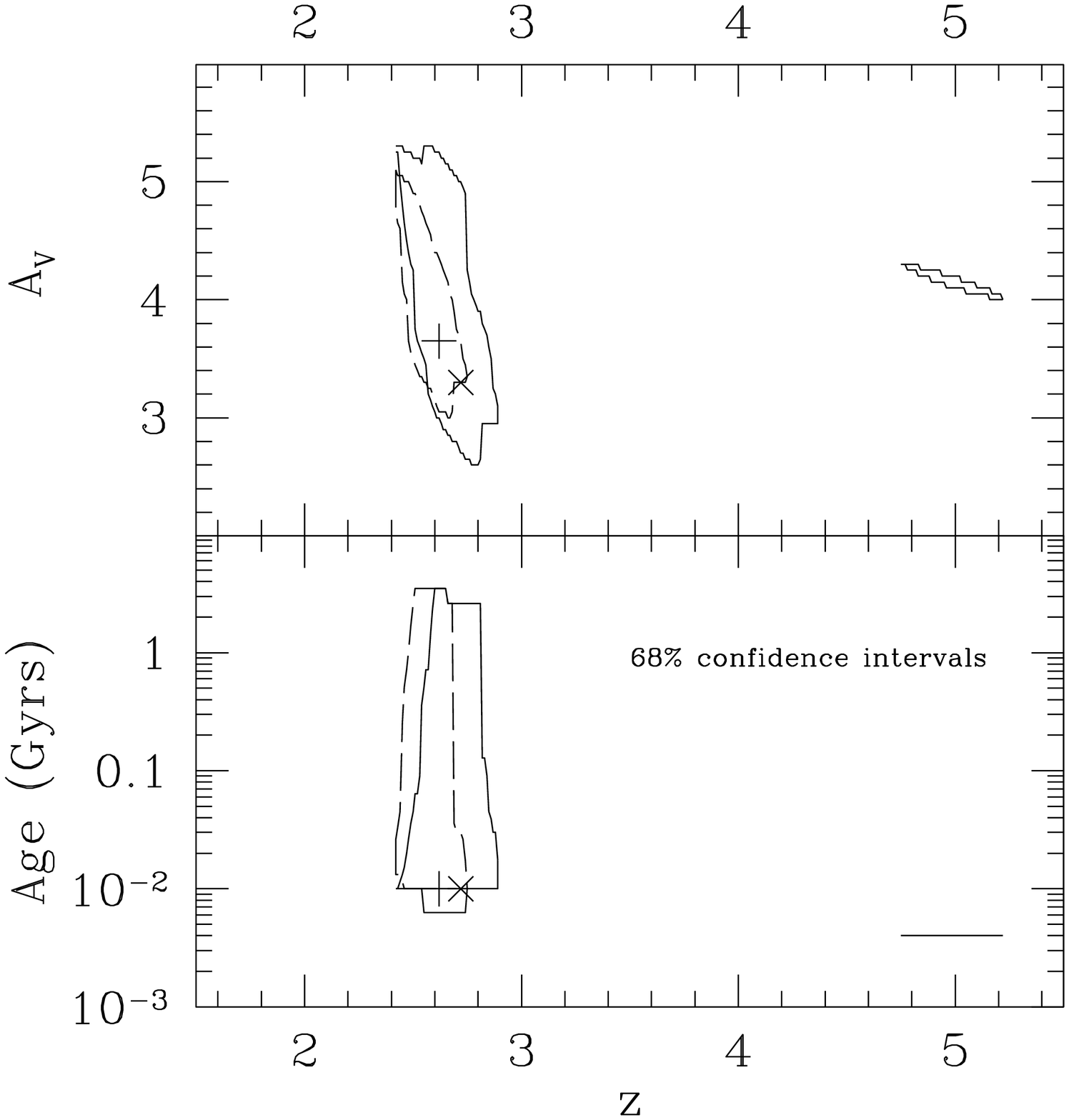}
\caption{The 68\% confidence intervals from \hyperz~ in the
redshift-extinction and redshift-age planes. The solid curves denote the
confidence intervals for image A; the dashed curves are for image B.  The
crosses (plus signs) denote the values corresponding to the minimum $\chi^2$
for image A (B). There is a small secondary peak in the solutions for image A
at $z=5$.\label{fig:photzerrors}}
\end{figure}

 Since these stellar templates do not include PAH emission, for the
photometric redshifts we fit only the data shortward of 10$\mu$m.  \hyperz~
yields a best fit redshift $z=2.72^{+0.19}_{-0.32}$ for image A (90\%
confidence; $\chi_\nu^2=0.5$).  A small secondary peak in the redshift
probability distribution is observed at $z=5$ (Fig. \ref{fig:photzerrors});
however, this redshift is implausible because of the high observed flux at
24$\mu$m.  An analysis of image B yields a similar redshift
$z=2.62^{+0.14}_{-0.22}$ ($\chi^2_\nu=0.97$) with no secondary peak.  In
\citet{bradac2006} we speculated that the source might be at $z>6$, but this
possiblity is now excluded at high confidence ($\Delta\chi^2>8$).  The best
fit spectral template (Fig. \ref{fig:sed}) corresponds to a dusty starburst
galaxy with $A_V=3.3^{+2.2}_{-0.8}$ (90\%) and an age of $<30$ Myr. As can be
seen in the figure, the robustness of the photometric redshift is largely due
to the fact that the \irac~data span the 1.6$\mu$m bump at this redshift.

At the best-fit redshift the $7.7\mu$m and 8.6$\mu$m PAH features (blended in
this spectrum) are redshifted beyond the 24$\mu$m window, in which case the
observed 24$\mu$m emission is dominated by the 6.24$\mu$m PAH feature.  In a
concurrent program \citep{wilson2008a} have also identified this galaxy as a
bright millimeter source, and obtain a consistent redshift ($z=2.7$) via an
empirical relation for SMGs between redshift and \irac~colors.

\subsection{Magnification and Additional Images}
\label{sec:magnification}
The magnification map was obtained from the strong (information from multiply
imaged systems) and weak (measuring distortion of background galaxies)
gravitational lensing data. It is the same reconstruction as in
\citet{bradac2006}. It is performed on a pixelized grid and does {\it not}
assume a specific form of the underlying gravitational potential.  At the
location of image A we compute a magnification $|\mu| \sim 25$ and at image B
we obtain $|\mu| \sim 50$ for a galaxy at $z=2.7$. The critical curve (points
of infinite magnification) passes between images A and B (see
Fig.\ref{fig:bigimg}) -- their parity is reversed -- supporting the hypothesis
that A and B are indeed multiple images of the same source. These
magnifications change by $<$20\% within the range of allowed photometric
redshifts (90\% confidence interval, see \S\ref{sec:sed}).  The measured flux
ratio of the two images is $1.47\pm0.05$ which is simlar to the flux ratio of
$\sim 2$ given by our lens model.  In the subsequent discussion we will quote
the stellar mass and star formation rate in terms of $(\mu_A/25)^{-1}$ to
reflect the inherent magnification uncertainty.

Given both lensing model and positions of images A and B, we can search for
additional images of this source.  We identified image C (Figure
\ref{fig:bigimg}) at $(\alpha_{2000},\delta_{2000})=$(06:58:33.4,-55:57:29)
using the same initial lens model, which does not include information from
images A and B of this object.  The photometry for image C, described in \S
\ref{sec:data}, indicates that this image is a factor of $2.2$ fainter than
image A, while our lens model predicts a factor of four. Given the
uncertainties these images are still consistent with being multiple images.
The best-fit photometric redshift derived for this source is
$z=2.82^{+0.18}_{-0.20}$.

\subsection{Stellar Mass}

To estimate the stellar mass we use the code \kcorrect~\citep{blanton2007},
which fits a linear combination of \citet{bc03} templates based upon Padova
1994 isochrones and spanning a range in metallicity ($0.005-2.5$ times solar)
and age (1 Myr to 13.75 Gyr). The required inputs are the photometric redshift
from \hyperz~ and the \irac~photometry, reddening corrected using the $A_V$
from \hyperz~ and the \citet{calzetti2000} reddening law. In this analysis and
\S\ref{sec:sfr} we focus upon image A, but note that equivalent results hold
for image B.  For $z=2.72$ and $A_V=3.3$ we obtain a stellar mass
$M_*=1.5\times10^{10}(\mu_A/25)^{-1}$ M$_\odot$.\footnote{As noted by
\citet{maraston2006}, templates that include the contribution of TP-AGB stars
to the spectrum can yield stellar masses roughly a factor of two lower. These
templates are not currently implemented in \kcorrect.}  Thus, we find that
this lensed galaxy is massive -- similar in mass to LIRGs at lower redshift
\citep{caputi2006b}.

\subsection{Presence of an AGN}
\label{subsec:agn}

In the sections above we have determined the redshift and stellar mass
assuming that the observed SED is dominated by stars. It is true however that
dusty starbursts and active galactic nuclei are difficult to discriminate at
the source redshift \citep[e.g., see][]{barmby2006}.

The simplest discriminator between the two contributors is spatial extent --
any spatially extended emission must be stellar rather than due to an AGN.  A
visual inspection of Figure \ref{fig:images} demonstrates that the
\irac~images do exhibit a modest extension perpendicular to the critical
curve, arguing that the flux is not purely from an AGN.  Next, we consider the
\chandra~observations to search for evidence of AGN activity. We find that the
lensed source is a non-detection in our 500 ks exposure ($f<3.6\times10^{-16}$
erg s$^{-1}$ cm$^{-2}$ unabsorbed, $3\sigma$, for 0.5-2 keV).  Comparing with
the local X-ray to mid-IR relations of \citet{krabbe2001}, we find that a
local Seyfert galaxy of comparable mid-IR luminosity should be more than a
factor of 10 brighter than this limit, whereas non-detection is consistent
with the expected relation for a starburst galaxy.  While neither of the above
arguments exclude an additional contribution from a central AGN, they do argue
that we are not looking at a purely AGN spectrum.

Given the 24$\mu$m data, we can also consider whether this source, which is
also a millimeter galaxy \citep{wilson2008a}, has mid-IR colors consistent
with a starburst or AGN. Comparing with the distribution of 24$\mu$m-8$\mu$m
vs. $8\mu$m-4.5$\mu$m colors for SMGs in \citet{pope2008}, we find that the
designation is ambiguous. The source lies close to, but outside the regime
defined in \citet{pope2008} for starburst galaxies, arguing that the observed
SED may be a composite with AGN and starburst contributions.

\subsection{Star Formation Rate}
\label{sec:sfr}

The only means of estimating the star formation rate with the existing data is
via the strength of the PAH emission. There are two  main caveats to this
approach.  First, any AGN contribution will bias our estimate of the star
formation rate.  Second, there exists large scatter in the relation between
8$\mu$m emission and star formation as traced by other methods
\citep{calzetti2008}. 

Keeping the above caveats in mind, we cautiously proceed with deriving a rough
estimate of the star formation rate. To do so, we first convert the observed
24$\mu$m luminosity to $L_{IR}$ and then use the local \citet{kennicutt1998}
relation to convert $L_{IR}$ to star formation rate.  Using the templates and
code from \citet{chary2001} to fit the 24$\mu$m flux, assuming $z=2.7$, we
derive a best-fit $L_{IR}=5\times10^{11} (\mu_A/25)^{-1}$ L$_\odot$. The
corresponding implied star formation rate is $\mathrm{SFR}\sim 90
(\mu_A/25)^{-1} \mathrm{M}_\odot$ yr$^{-1}$.  As a consistency check, we also
derive the rest-frame 8$\mu$m luminosity and convert to $L_{IR}$ using
$L_{8\mu\mathrm{m}}/L_{\mathrm{IR}}=10$, consistent with recent results from
\citet{rigby2008} based upon a combination of local and $z\sim2$ galaxies. To
derive the rest-frame 8$\mu$m we use the same spectral index correction
($\alpha=2.296$) as in \citet{dey2008}. This approach yields a qualitatively
consistent total luminosity, $L_{IR}\sim 3\times10^{11}(\mu_A/25)^{-1}$
L$_\odot$.

In Figure \ref{fig:sed} we overlay the best-fit \citet{chary2001} template for
image B, redshifted to $z=2.7$, at wavelengths redward of 10$\mu$m. If the
magnification is a factor of two lower than our canonical value, which
bootstrap simulations indicate is the maximum by which we may be
overestimating $\mu_A$, this galaxy would lie at the borderline betwen LIRG
and ULIRG luminosity.  From their AzTEC millimeter data, \citet{wilson2008a}
also estimate $L_{IR}=10^{11}-10^{12}$ L$_\odot$ for this source.  The
estimated specific star formation rate for this galaxy, $SFR\approx 5$
Gyr$^{-1}$, is comparable to that of similar mass BM/BX galaxies at
$z=1.5-2.6$ \citep{reddy2006}.

\section{Discussion}

We have presented confirmation observations for a multiply imaged source
behind the Bullet Cluster, and identified a third, previously unknown image.
From our multiwavelength imaging we argue that the source is most consistent
with being a dusty ($A_V\sim3.3$), strongly star-forming galaxy at
$z\sim2.7$. At this redshift our mass model for the cluster core indicates
that the galaxy is highly magnified ($\mu_A\sim25$), implying a large
intrinsic stellar mass of $M_*\sim 2\times10^{10}$ M$_\odot$. We estimate a
star formation rate of $\sim90$ M$_\odot$ yr$^{-1}$ based upon the observed
flux in the 24$\mu$ band, which we assume to be dominated by emission from
redshifted PAH features.

The estimated intrinsic infrared luminosity of this galaxy ($5\times10^{11}$
L$_\odot$) qualifies it as a luminous infrared galaxy (LIRG), fainter than the
ULIRGs typically studied at this epoch.  The galaxy is also known to be an
exceptionally bright SMG \citep[13.5 mJy at 1.1 mm,][]{wilson2008a}, and thus
may be an ideal system for studying the connection between different classes
of infrared sources at lower intrinsic luminosity than has previously been
possible for this epoch. We anticipate that the galaxy will be detected in
scheduled deep \hst~NICMOS observations, providing information on the spatial
extent of the lensed images.

The Bullet Cluster, due to its large lensing cross-section, provides an
optimal environment in which to identify lensed galaxies such as the one
presented here.  However, even in this rare, massive cluster merger, we
detected only one lensed LIRG. This result highlights that the prospects are
not good for finding large samples of $z\sim2$, gravitationally-lensed
LIRGS. It is for this reason that each case must be highlighted and exploited.

\label{sec:conclusions}

\acknowledgements 
AHG thanks Ranga-Ram Chary, Arjun Dey, Jean-Paul Kneib, and Alexandra Pope,
and Grant Wilson for constructive discussions related to this work,
St\'{e}phane Charlot and Gustavo Bruzual for providing access to the Charlot
\& Bruzual 2007 models, and the anonymous referee for comments that significantly improved
the manuscript. The authors acknowledge support for this work from NASA/HST
grants HST-GO-10200, HST-GO-10863, and HST-GO-11099, as well as NASA/Spitzer
grant 1319141. MB also acknowledges support from NASA through Hubble
Fellowship grant \# HST-HF-01206.01 awarded by the Space Telescope Science
Institute.

{\it Facilities:} HST (ACS), Spitzer (IRAC,MIPS), CXO (ACIS), Magellan:Baade (PANIC)
\bibliographystyle{apj}
\bibliography{ms}

\end{document}